\documentclass[aps,prl,twocolumn,showpacs,superscriptaddress,groupedaddress]{revtex4}
\usepackage{amsmath,amssymb}
\usepackage{graphicx}% Include figure files
\usepackage{dcolumn}% Align table columns on decimal point
\usepackage{bm}% bold math
\usepackage{subfigure}
\usepackage{color}
\newcommand{\ber}{\begin{eqnarray}}
\newcommand{\eer}{\end{eqnarray}}
\newcommand{\bea}{\begin{equation}}
\newcommand{\eea}{\end{equation}}
\DeclareMathOperator\erf{erf}

\begin{document}

\title{Generalization of Stokes-Einstein relation to coordinate dependent damping and diffusivity: An apparent conflict}

\author{A. Bhattacharyay}
\email{a.bhattacharyay@iiserpune.ac.in}
\affiliation{Indian Institute of Science Education and Research, Pune, India}

\begin{abstract}
Brownian motion with coordinate dependent damping and diffusivity is ubiquitous. Understanding equilibrium of a Brownian particle with coordinate dependent diffusion and damping is a contentious area. In this paper, we present an alternative approach based on already established methods to this problem. We solve for the equilibrium distribution of the over-damped dynamics using Kramers-Moyal expansion. We compare this with the over-damped limit of the generalized Maxwell-Boltzmann distribution. We show that the equipartition of energy helps recover the Stokes-Einstein relation at constant diffusivity and damping of the homogeneous space. However, we also show that, there exists no homogeneous limit of coordinate dependent diffusivity and damping with respect to the applicability of Stokes-Einstein relation when it does not hold locally. In the other scenario where the Stokes-Einstein relation holds locally, one needs to impose a restriction on the local maximum velocity of the Brownian particle to make the modified Maxwell-Boltzmann distribution coincide with the modified Boltzmann distribution in the over-damped limit.
\end{abstract}
\pacs{05.40.Jc, 05.10,Gg, 05.70.-a}% PACS, the Physics and Astronomy
                             % Classification Scheme.
\keywords{Diffusion, Brownian motion, Coordinate dependent damping, Equilibrium, Stokes-Einstein relation.}%Use showkeys class option if keyword
\maketitle
\section{introduction}
Diffusion shows a lot of variety. Normal Fickian diffusion is characterized by a mean square displacement (MSD) which scales linearly with time and the corresponding probability distribution of position is Gaussian. Deviation of the MSD from this linear scaling with time is termed as anomalous diffusion which falls into super- or sub-diffusive category based on MSD scaling as $t^\alpha$ with $1<\alpha <2$ and $0<\alpha<1$ respectively. There are observed variant of diffusion which are normal according to the linear scaling of the MSD with time, but, are characterized by non-Gaussian distributions at intermediate times which crosses over to Gaussian at larger times \cite{wang2, wang1}. An averaging over a distributed diffusivity of particles of the system has been seen to result in the required Laplace form of the density distribution \cite{wang2, hapc}. This method is similar in spirit to the concept of superstatistics which employs a superposition of Boltzmann statistics at smaller scales to get a non-Boltzmann distribution with a variable intensive quantity at larger scales \cite{beck}. Chubinsky and Slater came up with the idea of diffusing diffusion (time dependent diffusivity) and by employing that they got a short time probability distribution of Laplace form which crosses over to a Gaussian form at large times \cite{chub}. Chechkin et al., developed a minimal model using the concept of diffusing diffusion embeded in a two component Langevin dynamics to capture this crossover between Laplace to Gaussian regime as well \cite{chec}.

Coordinate dependence of diffusivity and damping \cite{bere} of a Brownian particle (BP) is observed (or invoked) in many experiments \cite{fauc,lanc,volp,wolf} where the BP resides near a wall or a boundary. Position dependent diffusion is supposed to be playing major role in protein folding  \cite{best,humm,oliv}. The same is also invoked in hydrodynamic (large wave length) models of some optical systems \cite{yami,pauf} and open quantum systems \cite{sarg}. A covariant formulation of state dependent diffusion and related issues with equilibration in such systems has been reported by Polettini \cite{pole}. The difficulty of experimentally determining position dependent diffusivity in protein folding is highlighted in a recent paper by Foster et al., \cite{fost}.

It is generally believed that the hydrodynamic effects near a wall are at the origin of the coordinate dependence of diffusivity and damping of a BP \cite{fauc} and there may be other reasons as well. Imagine a BP diffusing in a finite space filled with some network of static obstacles. The BP will be subjected to a coordinate dependent diffusivity and damping almost everywhere in such a stationary crowded space. Another example could be the Brownian motion of a polymer or a protein in its state space which is finite. Depending upon relative proximity of the monomers or residues in various configurations the diffusivity and damping could become a function of state space. Due to the finite extent of the space and static inhomogeneity, when kept at a constant temperature, such a Brownian motion should equilibrate at large times. The purpose of this paper is to ask if the equilibrium distribution and other features of such a system is different from that when diffusivity and damping are constant.

In the present kind of a problem, while looking for equilibrium, we are basically dealing with quenched coordinate dependent diffusivity and damping which depend on bath degrees of freedom and also things other than bath degrees of freedom. Had this not been the case, i.e. if the finite space of Brownian motion is homogeneous (being characterized by a constant diffusivity and damping), we would be in a regime of equilibrium governed by the Stokes-Einstein relation (fluctuation-dissipation relation (FDR)). The theory of equilibrium Brownian motion is well established for such homogeneous spaces. One may reasonably ask - what happens in the general case? does the Stock's-Einstein relation get generalized or it gets modified?

An over-damped BP under confinement equilibrates with heat-bath at large times. The Boltzmann distribution (BD) characterizes position distribution of the BP in equilibrium when the damping and the diffusivity are constant. This is a well known and tested result. What happens when the diffusivity and the damping are functions of space (i.e., coordinate dependent) is a question people have pondered over a long time and there exists controversy \cite{soko,tupp}. The main theme of the approach to this problem has so far been to demand the BD as an irrevocable condition for equilibrium \cite{san1,lau,san2,mark,far1,far2}. This necessitates replacing constant diffusivity D and constant damping $\Gamma$ with coordinate dependent $\text D(x)$ and $\Gamma(x)$ (for example, in 1D) to generalize the BD of such systems to $\text P(x) = \text N e^{\frac{-\text V(x)}{\text D(x)\Gamma(x)}}=\text Ne^{\frac{-\text V(x)}{\text {kT}}}$ where N is a normalization constant, $\text V(x)$ is the potential that confines the particle, k is the Boltzmann constant and T is the temperature of the system. This indicates a local generalization of the Stokes-Einstein relation $\text D\Gamma = \text {kT}$ to $\text D(x)\Gamma(x) = \text {kT} $.

One of the central issues, here, would be whether or not to take $\text D(x)\Gamma(x) = \text {kT} $ as the local generalization of the Stokes-Einstein relation and in this paper we will look at two cases where $\text D(x)\Gamma(x) \neq \text {kT} $ and $\text D(x)\Gamma(x) = \text {kT} $ within the framework of Kramers-Moyal expansion without {\it a priori} imposition of BD as an equilibrium condition. There are other issues when one imposes BD for equilibrium in such systems, like: (a) In the derivation of the probability distribution using Smoluchowski equation one needs to keep the Fick's law in its constant diffusivity form giving the diffusion current density $j_{diff} = -\text D(x)\frac{\partial \text P(x)}{\partial x}$ \cite{lau}. (b) The BD does not include coordinate dependent diffusivity $\text D(x)$ or damping $\Gamma(x)$ and, thus, does not reflect the inhomogeneity of space which cannot be accommodated in a potential. We will see in what follows that, the clue to have consistent solution to this problem lies with taking the correct form of Fick's law over inhomogeneous space where the diffusivity is a function of coordinates.

We show in the first part of our results where $\text D(x)\Gamma(x) \neq \text {kT} $ that, all the above mentioned issues get naturally resolved if one goes by the analysis based on the Kramers-Moyal expansion (in the over-damped regime) and the resolution happens in an unexpected way. By deriving the Smoluchowski equation for such a BP using Kramers-Moyal  expansion, one gets the modification of the Fick's law to $j_{diff}=-\frac{\partial}{\partial x}{\text D(x)}\text P(x)$ instead of a generalization to $j_{diff}=-{\text D(x)}\frac{\partial}{\partial x}\text P(x)$. The Stokes-Einstein relation in this inhomogeneous case results from the equipartition of kinetic energy as $\langle \text D(x)\Gamma(x)\rangle = \text {kT}$ indicating that $\text D\Gamma = \text {kT}$ is only strictly valid for the constant diffusivity and damping. 

There is no controversy in deriving the Fokker-Planck equation of a generalized Langevin dynamics that includes inertial term. This is so because one can easily convert the problem to a stochastic dynamics with additive noise. The distribution one gets here is a direct generalization of the Maxwell-Boltzmann (M-B) form. Moreover, correspondence between this generalized M-B distribution at the over-damped limit to the modified BD as obtained from the Smoluchowski dynamics produces ${\text D(x) = C\Gamma(x)}$ where $C = \frac{\langle{\text D(x)}^2\rangle}{{\text {kT}}}=\frac{\text {kT}}{\langle \Gamma(x)^2 \rangle}$ is a constant in equilibrium. 

It can be easily inferred that in the presence of the proportional relationship ${\text D(x) = C\Gamma(x)}$ between the local diffusivity and damping the product $\text D(x)\Gamma(x)$ cannot be locally equivalent to any quantity of an independent physical origin because that will make diffusivity locally inversely proportional to damping in direct conflict with ${\text D(x) = C\Gamma(x)}$. This subtle constraint will set the impossibility of having local temperatures in the form of $(\text D(x)\Gamma(x))/\text k$ in equilibrium of such systems. As a result $\text D(x)\Gamma(x)$ comes out to be the local energy scale that sets the width of distributions and this is only equal to the thermal energy scale set by the bath on an average over the inhomogeneity.

In this part, we will see that the general theory in the presence of equipartition of energy gives Stocks-Einstein relation in homogeneous space (constant D and $\Gamma$), but, the homogeneous limit of the relation in weakly inhomogeneous space does not exist. This observation will turn out to be crucial to rule out the extension of the theory for homogeneous space to the case of even weakly non-homogeneous cases when $\text D(x)\Gamma(x) \neq \text {kT} $. The non-existence of this homogeneous limit indicates a severe constraint on the Stokes-Einstein relation and its use even in weakly inhomogeneous space.

In the second part, we go by local validity of the Stokes-Einstein relation within the realm of an analysis strictly based on Kramers-Moyal expansion. We show that a crucial consideration is needed to make the two over-damped limits - one taken on the dynamics and the other taken on the modified M-B distribution - coincide. The consideration is that the local normalization of the velocity distribution cannot be done on limits from $-\infty$ to $+\infty$. The velocity limits on the integral has to be set between the quantities $-\text D(x)/\text L$ and $+\text D(x)/\text L$ where $\text L$ is the only length scale available that does not involve diffusivity and this length scale is the system size. Note that the Stokes-Einstein relation holds locally, $\text D(x) = \text {kT}/\Gamma(x)$. This means, the local maximal velocity limit is inversely proportional to 
$\Gamma(x)$ and is proportional to thermal energy $\text {kT}$ given a system size $\text L$ which makes sense. This we identify as an important requirement for the Stokes-Einstein relation to hold locally when diffusivity is coordinate dependent and the over-damped limit on modified M-B distribution resulting in the required modified BD that comes from the Smoluchowski equation. 

The Kramers-Moyal expansion is a formal procedure perfectly suited for a coordinate dependent diffusivity. The equilibrium distribution that results from the Smoluchowski equation as obtained from the Kramers-Moyal expansion is a modified Boltzmann distribution with the diffusivity $\text D(x)$ dependent amplitude \cite{ari1}. This is an expected equilibrium distribution as compared to the BD because the BD does not manifest the broken spatial homogeneity. Thus, beyond the so far used methods of essentially extending the Brownian motion theory of a homogeneous space to inhomogeneous conditions, if one follows already existing method (Kramers-Moyal expansion) for inhomogeneous space, one gets a set of consistent results and possibly the clue as to why the notions belonging to the homogeneous space theory cannot be extended to inhomogeneous situations. The present analysis based on Kramers-Moyal expansion indicates either the Stokes-Einstein relation is not locally applicable or when it holds even locally there exists a maximum limit of locally accessible velocity of the BP at each point in space.

The plan of the paper is as in the following. We first consider the over-damped Brownian dynamics and employ the Kramers-Moyal expansion to find out the Smoluchowski equation and its equilibrium solution as the modified BD. We then derive the Fokker-Planck dynamics for the generalized Langevin equation of the system which includes the inertial term. Following that we show our results in two subsections. In one subsection we employ equipartition to recover Stokes-Einstein relation in homogeneous space. We then take the over-damped limit of the generalized M-B distribution to compare this with the modified BD to get the relation between the $\text D(x)$ and $\Gamma(x)$. In the next subsection we consider the local validity of the Stokes-Einstein relation $\text D(x)\Gamma(x) = \text {kT} $ and show how one has to modify the integration limits of the local velocity normalization to get to the over-damped limit that follows from the Smoluchowski dynamics. We conclude the paper with a discussion of main results.  

\section{over-damped dynamics} 

Let us consider a 1D model of Brownian motion (for the sake of simplicity) as
\ber\nonumber
\dot{x} &=& v\\
m\dot{v} &=& -m\zeta(x)v+ {\text F(x)} + m\zeta(x)\sqrt{2{\text D(x)}}\eta(t), 
\eer
where $x$ is the position of the BP and $v$ is its velocity. We have kept the mass $m$ of the BP explicitly present for the ease of taking the over-damped limit, $m\zeta(x) = \Gamma(x)$ is the damping coefficient and F($x$) is an external force resulting from some potential ${\text F(x)} = -\frac{d{\text V(x)}}{dx}$. The Gaussian white noise of unit strength is represented by $\eta(t)$.

At the over-damped limit of this dynamics we get the very standard form of the  equation
\bea
\dot{x} = \frac{\text F(x)}{\Gamma(x)} + \sqrt{2{\text D(x)}}\eta(t).
\eea

In what follows, we will never impose any {\it a priori} relationship between the D($x$) and $\Gamma(x)$. The relations will follow from the over-damped limit of the generalized M-B distribution and the equipartition of kinetic energy. Let us have a look at a few well known but important details of the over-damped Langevin dynamics eqn.(2). In the absence of the force $\text F(x)$ it represents free diffusion. The diffusivity $\text D(x)$ gets defined by the dynamics in the presence of the Gaussian noise $\eta(t)$. Thus, ${\text D(x)} = \frac{\langle (x(t+\delta t)-x(t))^2\rangle}{2\delta t}$ where $x\equiv x(t)$ \cite{ari1}. 

The diffusion time scale $\delta t$ would depend on the locality of the D($x$) and the average is over noise. Inclusion of the force term F($x$) makes the damping $\Gamma(x)$ explicitly appear and fix the local drift current. We, therefore, are effectively considering normal diffusion here, the only modification is in the local character of the diffusivity and the damping. A very important property of normal diffusion is the isotropy of the process and in the present case although the diffusivity is inhomogeneous in space, it is isotropic, i.e., the same in both directions at every point in one dimensional space.

Let us first have a look at the Smoluchowski equation for the over-damped dynamics (eqn.(2)) using Kramers-Moyal expansion \cite{ari1}. The Kramers-Moyal expansion gives dynamics of probability density P($x,t$) as
\bea
\frac{\partial\text P(x,t)}{\partial t} = \sum_{n=1}^{\infty}{\left (-\frac{\partial}{\partial x}\right )^nD^{(n)}(x,t)\text P(x,t)},
\eea
where the expansion coefficients are 
\bea
D^{(n)}(x,t) = \frac{1}{n\,!}\lim_{\tau\to 0}\frac{1}{\tau}\langle[\xi(t+\tau)-x]^n\rangle
\eea
with $\xi(t)=x$ and the angular brackets indicate average over noise \cite{risk}. Consistent with Pawula's theorem, there will be two terms on the r.h.s., of the Smoluchowski equation for the BP whose dynamics is given by the Langevin equation (eqn.(2)) as 
\bea
\frac{\partial {\text P(x,t)}}{\partial t} = \frac{\partial}{\partial x}\left [-\frac{{\text F(x)}{\text P(x,t)}}{\Gamma(x)} + \frac{\partial\text D(x){\text P(x,t)}}{\partial x} \right ].
\eea

At this stage, a discussion on the so-called spurious current is in order. The drift current density to be $j_{drift}=\frac{\text F(x)}{\Gamma(x)}\text P(x)$ is determined by the drift velocity $v_d(x)=\frac{\text F(x)}{\Gamma(x)}$ which results from a balance between the damping term and the external force. In the over-damped limit this force balance is always there and does not depend on the coordinate dependence of damping. This is so because the over-damped limit is taken by setting $m\to 0$ and $\zeta(x) \to \infty$ such that $\Gamma(x)$ is finite. This limit practically sets the relaxation time of the system $\tau(x) = \frac{1}{\zeta(x)} \to 0$ everywhere and thus the relaxation time scale becomes negligible compared to the diffusion time scale $\delta t$. 

Due to non-validity of mean value theorem on stochastic integrals with multiplicative noise, a convention is needed to evaluate the drift coefficient $D^{(1)}(x,t)$ and here comes the It\^o vs Stratonovich dilemma \cite{ito,stra}. Where It\^o convention correctly gives the drift current in its form, the Stratonovich convention produces spurious drift current on top of it which has to be neglected in a straight forward manner if one wants strictly to be in the over-damped limit. The drift velocity is completely defined in the over-damped limit everywhere because of the reason that diffusion being isotropic at all points in space even when diffusivity is coordinate dependent the diffusion gradient cannot result in a drift current. It is essential to break isotropy of space to get a drift current, however, the coordinate dependent diffusivity does not do that symmetry breaking. The inhomogeneity of space due to $\text D(x)$ shows an apparent breaking of isotropy by the presence of gradients, but, diffusive transport remaining isotropic there cannot be a drift proportional to these gradients. If such a drift appears that appears as an artifact of the convention followed. Thus, it is not at all difficult to identify the spurious convention dependent component of drift current here. 

Moreover, the diffusion current does not involve any spurious contribution in any convention and, therefore, cannot be altered. This is exactly where, in the existing literature, manipulations are made. One not only cancels the spurious drift current but also throws away the part of the diffusion current appearing in the form $-\text P(x)\frac{d\text D(x)}{dx}$ to ensure Boltzmann distribution. For example, in the paper by Lau and Lubensky \cite{lau}, which explains the existing practice in this regard in a general way, one can identify the omission of the above mentioned part of the diffusion current in the considered definition of the diffusion current density as $J(x,t) = -{\text D(x)}\frac{\partial {\text P(x,t)}}{\partial x}$. But, how could this be done even when there is no spurious contribution in diffusion current? As is clearly mentioned by Lau and Lubensky \cite{lau}, this is done to get the BD as the equilibrium solution of the resulting Smoluchowski equation. 
  
The equilibrium distribution that results from the Smoluchowski dynamics as given by eqn.(5) is 
\bea
\text P(x) = \frac{\text N}{\text D(x)}\exp{\int_{-\infty}^x \frac{\text F(x^\prime)}{\text D(x^\prime)\Gamma(x^\prime)}\text dx^\prime}.
\eea
This is a modified Boltzmann distribution in the presence of coordinate dependent diffusivity and damping where $\text N = \left [\int_{-\infty}^{\infty}{\frac{\text dx}{\text D(x)}}\exp{\int_{-\infty}^x \frac{\text F(x^\prime)}{\text D(x^\prime)\Gamma(x^\prime)}dx^\prime} \right ]^{-1}$ is a normalization constant. Note that, the temperature of the bath does not show up in this expression since we have not yet considered any relation between the diffusivity and damping as the one results from the Stokes-Einstein relation in homogeneous space. Obviously, we do not want to impose the Stokes-Einstein relation. The way to bring in the temperature is to employ the equipartition of kinetic energy of the BP and for that we will be needing to find out the equilibrium distribution for the model involving inertial term given by eqs.(1).

The modified Boltzmann distribution as shown in eqn.(6) can always be given a Boltzmann form by exponentiating the D($x$) dependent amplitude. This would result in an effective potential involving the D(x) and $\Gamma(x)$ as shown in \cite{ari1}. Making use of this effective potential, one can write a Langevin dynamics with additive noise to simulate equilibrium fluctuations of a system. After all, it is the equilibrium fluctuations of the Langevin dynamics which are of any practical use. If the temperature is identified properly, then this alternative procedure can possibly work fine for a whole class of stochastic problems in inhomogeneous space.

Before going to the next section to capture the temperature in this formalism let us have critical look at the issue as to why the Boltzmann distribution cannot be an acceptable equilibrium distribution for the over-damped dynamics as given by eqn.(2) whereas the modified Boltzmann distribution is a perfectly acceptable equilibrium distribution. It is important to notice that, if the eqn.(2) is characterized by a Boltzmann distribution in equilibrium the average mean velocity $\langle \dot x \rangle = \langle \frac{\text F(x)}{\Gamma(x)} \rangle$ is not identically equal to zero whereas $\langle \frac{\text F(x)}{\Gamma(x)} \rangle =\int_\infty^\infty{dx\frac{\text F(x)}{\text D(x)\Gamma(x)}\text P(x)}=\int_0^0{d\text P(x)} \equiv 0$ when $\text P(x)$ is the modified Boltzmann distribution without the $1/D(x)$ normalization factor as is given by eqn.(6). This is a crucial check. The equilibrium distribution is stationary by construction as the BP equilibrates at the minimum of a potential. Existence of this average current due to Boltzmann distribution will produce entropy in contradiction with the thermodynamic demand of equilibrium to be the highest entropy state under given conditions. Had the manipulations normally done on the Smoluchowski dynamics to get the Boltzmann distribution for eqn.(2) been correct this inconsistency would have not resulted. However, the appearance of this inconsistency clearly indicates that the modified Boltzmann distribution as obtained from the methods following the Kramers-Moyal expansion is just perfectly consistent to be the equilibrium distribution.   

\section{Generalized Langevin dynamics}

Considering the change of variable $u = \frac{v}{\chi(x)}$ where $\chi(x)=\zeta(x)\sqrt{2{\text D(x)}}$, eqs.(1) take the form

\ber\nonumber
\dot{x} &=& \chi(x)u\\
\dot{u} &=& -\zeta(x)u - \chi^\prime(x) u^2 + \frac{{\text F(x)}}{m\chi(x)} + \eta(t).
\eer

In eqs.(7), the $\chi^\prime(x)=\frac{\partial \chi(x)}{\partial x}$ and, these equations being in additive noise form, its Fokker-Planck dynamics can be derived in a straight forward manner. The Fokker-Planck dynamics for eqn.(7) is
\begin{widetext}
\bea
\frac{\partial \text P(x,u,t)}{\partial t} = -\frac{\partial}{\partial x}\chi(x)u{\text P(x,u,t)} - \frac{\partial}{\partial u}\left [-\zeta(x)u - \chi^\prime(x) u^2 + \frac{{\text F(x)}}{m\chi(x)} \right ]{\text P(x,u,t)}+ \frac{1}{2}\frac{\partial^2}{\partial u^2}{\text P(x,u,t)}.
\eea
\end{widetext}

To obtain the stationary equilibrium distribution with the detailed balance maintained, one sets the part of the equation involving the operators symmetric in $u$ to zero to obtain the velocity distribution. This requirement of detailed balance in equilibrium (see for reference chap. 6 of \cite{gard}) requires the r.h.s., of eqn.(8) be separated in the following manner for a stationary solution.

\begin{widetext}
\bea
-\frac{\partial}{\partial x}\chi(x)u{\text P(x,u)} + \frac{\partial}{\partial u}\chi^\prime(x)u^2{\text P(x,u)}-\frac{\partial}{\partial u}\frac{{\text F(x)}}{m\chi(x)}{\text P(x,u)}= -\frac{\partial}{\partial u}\left [\zeta(x)u{\text P(x,u)} + \frac{1}{2}\frac{\partial}{\partial u}{\text P(x,u)}\right ].
\eea
\end{widetext}
Setting the current density within the square bracket on the r.h.s., of the above equation to zero one gets the Maxwellian distribution of the velocity and the stationary probability density now assumes the shape 
\bea
{\text P(x,u)} = \text P(x)\text M(x)e^{-\zeta(x)u^2} = \text P(x)\text M(x)e^{-\frac{\zeta(x)v^2}{\chi(x)^2}}.
\eea 
In the above mentioned expression for probability density the local normalization factor $\text M(x) = \frac{1}{\chi(x)}\sqrt{\frac{\zeta(x)}{\pi}}$ for the velocity ($v$) distribution is explicitly considered. With these, eqn.(9) now takes the form
\begin{widetext}
\ber\nonumber
-u\frac{\partial}{\partial x}\chi(x){\text P(x)}{\text M(x)}e^{-\zeta(x)u^2} &+& 2u\chi^\prime(x){\text P(x)}{\text M(x)}e^{-\zeta(x)u^2}-2u^3\chi^\prime(x)\zeta(x){\text P(x)}{\text M(x)}e^{-\zeta(x)u^2}\\ &+&\frac{2u\zeta(x){\text F(x)}}{m\chi(x)}{\text P(x)}{\text M(x)}e^{-\zeta(x)u^2} = 0.
\eer
\end{widetext}
Removing the common factor of $u$ from all the terms and then integrating out $v$ while keeping in mind that the average $\langle v^2 \rangle_{local} = \frac{\chi(x)^2}{2\zeta(x)}$ we get
%\begin{widetext}
\ber\nonumber
-{\text P(x)}\chi^\prime(x) &-& \chi(x)\frac{\partial}{\partial x}{\text P(x)} + 2{\text P(x)}\chi^\prime(x) - {\text P(x)}\chi^\prime(x)\\ &+&\frac{2\zeta(x){\text F(x)}}{m\chi(x)}{\text P(x)}=0.
\eer
%\end{widetext}
Eqn.(12) results in a distribution over position space
\bea
{\text P(x)} = e^{\int_{-\infty}^xdx^\prime\frac{2\zeta(x^\prime){\text F(x^\prime)}}{m\chi(x^\prime)^2}}.
\eea
Including all the terms, therefore, the generalized M-B distribution is
\bea
P(x,v)=N\sqrt{\frac{m}{2\pi\Gamma(x)\text D(x)}}e^{\int_{-\infty}^xdx^\prime\frac{{\text F(x^\prime)}}{\Gamma(x^\prime)\text D(x^\prime)}}e^{-\frac{mv^2}{2\Gamma(x)\text D(x)}}
\eea
where $N = \left [\int_{-\infty}^{\infty}{\text dx \text P(x)}\right ]^{-1}$ is an overall normalization constant. Note that, this M-B distribution is an exact generalization of the M-B distribution over homogeneous space \cite{swab}. If one replaces $\text D(x)$ by $\text D$ and $\Gamma(x)$ by $\Gamma$, one would get the standard M-B distribution of a BP over homogeneous space.
\subsection{Stokes-Einstein relation does not hold locally}
So far, the temperature has not been introduced in the expressions we have got and that now can easily be obtained from the equipartition of kinetic energy. The equipartition of energy is a general feature of equilibrium and the kinetic energy being quadratic in momentum its average value is $\frac{{\text k \text T}}{2}$. The average must be done over the whole phase space. Using the general M-B distribution the equipartition results in  
\bea
\langle mv^2\rangle = N\int_{-\infty}^{\infty}{\text dx \text P(x) \frac{m\chi(x)^2}{2\zeta(x)}} = \langle \Gamma(x)\text D(x) \rangle = \text {kT}
\eea
where the angular brackets indicate a space average over the bounded region in which the BP has equilibrated with the bath. Obviously, when diffusivity and damping are constant, we recover the Stokes-Einstein relation $\text D=\frac{\text {kT}}{\Gamma}$ from the equipartition and this relation does not hold in general for a coordinate dependent damping and diffusion. 

Important to note that, arriving at the Stokes-Einstein relation at the homogeneous case justifies {\it a posteriori} the use of equipartition relation eqn.(15). Equipartition of kinetic energy giving $\frac{1}{2}{\text {kT}}$ is a consequence of the M-B distribution, however, the distribution we have arrived at is a generalized form without the temperature being explicitly present. The recovery of Stokes-Einstein relation for constant D and $\Gamma$ now raises the question - does the homogeneous limit exist where the Stokes-Einstein relation can be used at least for weak inhomogeneity? We will try to find an answer to this question in the following. 

The relation between the coordinate dependent damping and diffusion can be arrived at by taking the over-damped limit $m\to 0$ and $\zeta (x) \to \infty$ keeping $\Gamma (x)$ finite on the generalized M-B distribution (eqn.(14)) and comparing that with the modified BD as already obtained in eqn.(6). This limit sets the factor $e^{-\frac{mv^2}{2\Gamma(x)\text D(x)}}$ to unity and the resulting limit of the normalization factor $\sqrt{\frac{m}{2\pi\Gamma(x)\text D(x)}}\to 0$ is a consequence of the flatness of the velocity distribution however, the normalization factor must be kept explicitly present in the expression of the distribution. At this limit, correspondence between the generalized M-B distribution and the generalized BD needs $\text D(x) = C\Gamma(x)$ where the proportionality constant comes out from the equipartition to be $C = \frac{\langle{\text D(x)}^2\rangle}{{\text {kT}}}=\frac{\text {kT}}{\langle \Gamma(x)^2 \rangle}$.

These two relations (1) $\text D(x) = C\Gamma(x)$ and (2) $\text D=\frac{\text {kT}}{\Gamma}$ are completely consistent so long one takes into account the fact that the latter is valid strictly for constant diffusivity and damping. It is obvious that even at weak inhomogeneity limit the Stokes-Einstein relation cannot approximate for the relation $\text D(x) = C\Gamma(x)$. This fact can easily be checked by Taylor expanding two expressions for $\text D(x)$ namely (a) $\text D(x)=\frac{\text {kT}}{\Gamma(x)}$ and (b) $\text D(x) = \frac{\text {kT}}{\langle \Gamma(x)^2 \rangle}\Gamma(x)$ for small $\frac{d\Gamma(x)}{dx}$ which takes into account the weak spatial variation of $\Gamma(x)$ over its average value $\Gamma = \langle \Gamma(x) \rangle $. Consider the average of the diffusivity to be $\text D = \langle \text D(x) \rangle$.

Taking the relation (a) into account and truncating the Taylor expansion about $\langle \text D(x)\rangle = \text D$ and $\langle \Gamma(x)\rangle =\Gamma$ at the second term we get
\bea
\text D + \left (\frac{d\text D(x)}{d\Gamma(x)}\frac{d\Gamma(x)}{dx}\right )_{\text D,\Gamma}\delta x = \frac{\text {kT}}{\Gamma} - \frac{\text {kT}}{\Gamma^2}\left (\frac{d\Gamma(x)}{dx}\right )_{\text D,\Gamma}\delta x,
\eea
where the expansion is truncated at the second term due to smallness of $\frac{d\Gamma(x)}{dx}$ in the weakly inhomogeneous space. This gives
\bea
\left (\frac{d\text D(x)}{d\Gamma(x)}\right )_{\text D,\Gamma} = - \frac{\text {kT}}{\Gamma^2}.
\eea
Going by the relation (b) and the same procedure 
\bea
\left (\frac{d\text D(x)}{d\Gamma(x)}\right )_{\text D,\Gamma} = C = \frac{\text {kT}}{\Gamma^2}.
\eea
So, the mismatch of the sign cannot be cured unless $\text T\to 0$ or equivalently $\Gamma \to \infty$ which is essentially a non-stochastic limit. Therefore, the relations (a) and (b) cannot be limiting cases of each other at even small inhomogeneity. This indicates the Stokes-Einstein relation cannot be generalized to situations where the diffusivity and damping are even weakly coordinate dependent. 

Let us have a closer look at the implications of the non-existence of this limit. The equipartition gives a generalization of the Stokes-Einstein relation as $\langle \Gamma(x)\text D(x) \rangle = \text {kT}$ and the correspondence gives $\text D(x) = C\Gamma(x)$. While the equipartition clearly indicates that the temperature is defined globally, the relation $\text D(x) = C\Gamma(x)$, which is at conflict with the local generalization of the Stokes-Einstein relation, relates the local fluctuation and dissipation. Moreover, the latter indicates, the local temperature is proportional to $\frac{D(x)}{\Gamma(x)}$ and not to $\text D(x)\Gamma(x)$ as would be the demand of a generalized Stokes-Einstein relation.

The energy scale $\text W(x)=\text D(x)\Gamma(x)$ may be interpreted in analogy as $\text {kT}(x)$ over an inhomogeneous space, however, this analogy cannot bring in a local temperature $\text {kT}(x)$ of an independent physical origin (i.e. a property of the bath) than what the product $\text D(x)\Gamma(x)$ itself is. This is so because, existence of any other independent physics (for example thermal) giving rise to such a quantity will impose an inverse relation between the local diffusivity and damping in direct conflict with $\text D(x) = C\Gamma(x)$. Therefore, it is clear that, although an analogy apparently exists, but, it is of no physical consequence to actually create thermal gradients in equilibrium as captured by the modified Boltzmann distribution.  Thus, the appearance of the relation $\text D(x) = C\Gamma(x)$ preserves the basic tenet of existence of no temperature gradients in equilibrium. In other words, the failure of the local generalization of the Stokes-Einstein relation rids us of the problem of appearance of local temperature in the equilibrium scenario of such a spatially inhomogeneous space.
\subsection{Stokes-Einstein relation holds locally}
The knowledge gained in the previous subsection indicates that if Stokes-Einstein relation holds locally, the over-damped limit on the generalized Maxwell-Boltzmann distribution cannot correspond to the modified Boltzmann distribution that we have got from the Smoluchowski dynamics because $\Gamma(x)\text D(x)$ is a constant $\text{kT}$. There is a simple way out of this problem. Although we are used to integrating over all velocities while normalizing the velocity distribution, however, the situation at hands indicates that we cannot do that when $\text D(x)$ is coordinate dependent. 

The natural local velocity cut off for such a system can be taken as $\text D(x)/\text L$ where L is the system size i.e. the length scale of the space in which the BP equilibrates. There is no other length scale available in this system than this which does not depend on $\text D(x)$ when Stokes-Einstein equation is locally valid. When employed, this gives a local normalization factor for the velocity distribution in the following way.
\bea
\int_{-\frac{\text D(x)}{\text L}}^{\frac{\text D(x)}{\text L}}{dve^{-\frac{mv^2}{2\Gamma(x)\text D(x)}}} = \sqrt{\frac{2\text {kT}}{m}}\int_{-\frac{\text D(x)}{\text L}\sqrt{\frac{m}{2\text{kT}}}}^{\frac{\text D(x)}{\text L}\sqrt{\frac{m}{2\text{kT}}}}{dz e^{-z^2}},
\eea
where $z=\sqrt{\frac{m}{2\text{kT}}}v$.

Eqn.(19) readily gives the value of the integral to be

$$
\sqrt{\frac{2\pi\text {kT}}{m}} {{\erf}}(\frac{\text D(x)}{\text L}\sqrt{\frac{m}{2\text{kT}}}),
$$
which at the overdamped limit can be written simply as $\sqrt{\pi}\text D(x)/\text L$ and that gives the normalization constant proportional to $1/\text D(x)$. 

The modified Maxwell-Boltzmann distribution in this case becomes

\bea
P(x,v)=\frac{N}{{\sqrt{\frac{2\pi\text {kT}}{m}}{\erf}}(\frac{\text D(x)}{\text L}\sqrt{\frac{m}{2\text{kT}}})}e^{\int_{-\infty}^xdx^\prime\frac{{\text F(x^\prime)}}{\Gamma(x^\prime)\text D(x^\prime)}}e^{-\frac{mv^2}{2\Gamma(x)\text D(x)}}.
\eea
Now, taking the over-damped limit i.e. $m\to 0$ on this we get,

\bea
P(x,v)=\frac{N}{D(x)}\exp{\int_{-\infty}^xdx^\prime\frac{{\text F(x^\prime)}}{\Gamma(x^\prime)\text D(x^\prime)}},
\eea

where in the above relations $N$ stands for normalization constant. Note that, eqn.(21) is identical to eqn.(6) and we get to the same  expression for the modified Boltzmann distribution at the over-damped limit by taking the limit both ways - on the dynamics and on the modified M-B distribution. When Stokes-Einstein relation holds locally, it needs the local maximum velocity be restricted to $\text D(x)/\text L$ is the physics which goes very much contrary to the common sense that at $m \to 0$ all velocities should be allowed. However, this length scale remains hidden in the normalization constant and instead of this, any other constant emergent length scale would result in the same. Here experiments can possibly look for the existence of an emergent length scale which fixes the local maximum velocities.

\section{discussion}

In this paper we have looked at the problem of a Brownian particle moving in a finite space where its diffusivity and damping are stationary and are coordinate dependent. We have been investigating the equilibrium of such a finite system. Coordinate dependent diffusivity and damping makes the space inhomogeneous even in the absence of a force, however, the isotropy of the space remains intact at every point over space in this diffusive process. A global force may break the isotropy of the system and result in drift current but the diffusion does not do that.

In the over-damped limit our approach has been to consider the Smoluchowski equation as obtained from Kramers-Moyal expansion and solve it for equilibrium distribution without imposing any condition. On the other hand, we derived the generalized Maxwell-Boltzmann distribution for equilibrium of the system and then took the over-damped limit on it. On comparison of results obtained from the over-damped dynamics and over-damped limit of the generalized M-B distribution we see that there exists a proportional relation between coordinate dependent diffusivity and damping when the Stokes-Einstein relation does not hold locally. The equipartition of energy results in recovery of Stokes-Einstein relation for constant diffusivity and damping. However, in terms of validity of the Stokes-Einstein relation the limit of the inhomogeneous space going to the homogeneous space does not exist.

On the other hand, when we have taken into consideration the local validity of the Stokes-Einstein relation, we see that, we have to impose a local maximal velocity limit to the velocity distribution to recover the modified Boltzmann distribution of over-damped limit from the generalized M-B distribution. The generalized M-B distribution in this case is not a straight forward generalization of the M-B distribution with constant diffusivity and damping and involves an error function in the normalization factor. This is an interesting situation where the local maximal velocity a BP can take is proportional to $\text {kT}$ and inversely proportional to $\Gamma(x)$. The equipartition will hold in this case locally unlike where the Stokes-Einstein relation is not locally valid.

These modified equilibrium distributions and relations between the local diffusivity and damping could be checked within present experimental access. To the knowledge of the author, experiments so far have not particularly looked for such an inversion of the Stokes-Einetein relation or local maximum velocity of a BP. On the contrary, Stokes-Einstein relation has been extensively employed to get diffusivity from damping and vice versa even when the diffusivity and damping are space dependent. These new results, which are based on already established formal methods, if experimentally verified, can have far reaching consequence on our present understanding of equilibrium of such systems.

Let us try to understand why such an equilibrium analysis of the Brownian motion in inhomogeneous space is important. Consider the biophysical environment of a cell. This is a very crowded and confined environment and of course the processes are not happening strictly in equilibrium in the true thermodynamic sense. However, many of the processes are weakly non-equilibrium stochastic processes whose statistics to be mostly governed by equilibrium fluctuations. In other words, many processes fall in the linear response regime where the equilibrium distribution dictates the physics. This is exactly the reason we care about an otherwise idealized equilibrium conditions because the same physics applies to a plethora of phenomena in the weakly non-equilibrium regime. The importance of the present results lie in this wide area of applicability.

{\section{Acknowledgement} {I would like to acknowledge discussions with J. K. Bhattacharjee.}

\end{document}